# An End-to-End Performance Comparison of Seven Permissioned Blockchain Systems

Frank Christian Geyer
Munich University of Applied Sciences
Munich, Germany
frank@coconut.sh

Hans-Arno Jacobsen
University of Toronto
Toronto, Canada
jacobsen@eecg.toronto.edu

Ruben Mayer
University of Bayreuth
Bayreuth, Germany
ruben.mayer@uni-bayreuth.de

Peter Mandl
Munich University of Applied Sciences
Munich, Germany
peter.mandl@hm.edu

## ABSTRACT

The emergence of numerous blockchain solutions, offering innovative approaches to optimise performance, scalability, privacy, and governance, complicates performance analysis. Reasons for the difficulty of benchmarking blockchains include, for example, the high number of system parameters to configure and the effort to deploy a blockchain network. In addition, performance data, which mostly comes from system vendors, is often opaque. We provide a reproducible evaluation of the performance of seven permissioned blockchain systems across different parameter settings. We employ an end-to-end approach, where the clients sending the transactions are fully involved in the data collection approach. Our results underscore the unique characteristics and limitations of the systems we examined. Due to the insights given, our work forms the basis for continued research to optimise the performance of blockchain systems.

## 1 INTRODUCTION

Following the hype around public blockchains such as Bitcoin [50] and Ethereum [26], practitioners and academics found that such systems do not suit the needs of business users due to their lack of performance, privacy, and governance. Hence, they developed permissioned systems such as Hyperledger Fabric [6, 16], Corda [23, 43] and Quorum [12, 22].

Blockchains can be run on custom hardware, can be configured as needed, can enforce a permission concept to keep data private, and they can offer different architectures according to a specific use case. Given the vast number of options available, it's challenging to predict expected performance.Vendors often tout remarkable performance [8, 16, 18, 48, 55], but there is lack of a fair comparison of systems under comparable conditions and workloads. As the permissioned blockchain systems widely differ in their design, benchmarking them is difficult.

Whilst various publications for benchmarking multiple blockchains already exist (cf. Table 1), these publications do not fully satisfy our needs to enable a comprehensive performance evaluation. By comprehensive performance evaluation, we mean examining multiple blockchains using different architectures with custom workload scenarios, which includes using custom smart contracts and changing system parameters.

To close this gap, we are conducting a performance evaluation of seven systems with a specially developed benchmarking system based on selected load scenarios. The seven systems comprise Corda Open Source (OS) [23, 43], Corda Enterprise [23, 43], BitShares [3, 28] which is based on the Graphene blockchain framework [36], Hyperledger Fabric (Fabric) [6, 16], Quorum [12, 22], Hyperledger Sawtooth (Sawtooth) [7, 41] and Diem [5, 55].

After a comprehensive review of the vast blockchain landscape, which comprises a plethora of approaches, we strategically selected seven unique and popular blockchains. Our choices were influenced by the desire to represent a diverse set of use case scenarios and, to an extent, by the necessity to start with a manageable subset given the rarity of performance benchmarks in the field. Importantly, these selected blockchain systems are not just theoretical constructs but have been employed in various real-world scenarios, further emphasizing their relevance and significance in the broader blockchain arena. These platforms not only stand as representatives of the broader ecosystem but also provide a balanced blend of functionalities, consensus algorithms, and architectural variances.

**Table 1: Overview of publications using custom benchmarking and performance evaluation frameworks.**

| Name | Analyzed Blockchains | Custom Smart Contracts | End-to-end scenario to measure metrics | Comparison of system parameters influence |
|---|---|---|---|---|
| BBB [47] | Ethereum [26] | No. | No. | No. |
| BCTMark [49] | Ethereum, Fabric [6, 16] | Yes, incl. three sorting algorithms. | No. | No. |
| BlockBench [25, 38] | Ethereum, Parity [4], Fabric | Yes, multiple, incl. SmallBank [14]. | No. | No. |
| Diablo [24, 39] | Algorand [29], Ethereum, Avalanche [52], Diem [5, 55], Quorum [12, 22], Redbelly [10], Solana [15] | Yes, 6 smart contracts. | No. | No. |
| DLPS [51] | Ethereum, Parity, Fabric, Hyperledger Indy [1], Quorum, Sawtooth [7, 41] | Yes, KeyValue-Write smart contract. | No. | No. |
| Gromit [37, 45] | Ethereum, Algorand, BitShares [3, 28], Diem, Stellar [42], Fabric, Avalanche | No. | No. | No. |
| xBCBench [53] | Ethereum, Fabric, Sawtooth, FISCO-BCOS [11] | Yes, different smart contracts. | No. | No. |
| **Our paper** | **Corda OS & Enterprise [23, 43], BitShares, Fabric, Quorum, Sawtooth, Diem** | **Yes, DoNothing, KeyValue-storage, banking app.** | **Yes, measurements by clients.** | **Yes, comparing seven systems.** |

Additionally, we examine the influence of the impact of network latency and the adjustment of the size of the network on the performance leading to a total of more than 3,000 executed experiments for the selected blockchain systems. To date and to the best of our knowledge, our paper is the only one paying attention to all relevant factors outlined in Table 1.
This paper makes the following contributions:

(1) We conduct an analysis of the performance of the blockchain systems using the results from the experiments executed by us. In this paper, we concentrate on two key metrics, which are the throughput in mean transactions per second and the mean finalization latency. Compared to existing work, our key metrics are obtained in an end-to-end scenario from a client perspective. Our contribution and the lessons learned can be used by researchers and practitioners to further understand and thus optimize the performance of blockchain systems.
(2) We use our results to show how each system is influenced by different parameter settings and highlight the most important key findings, which include that various systems cease their operation under high load and performance drops in different scenarios. These findings should create awareness for performance pitfalls when using these systems.
(3) We introduce the novel benchmarking framework COCONUT (an automatiC blOckChain perfOrmaNce evalUation sysTem). This open-source framework allows to fully reproduce our executed benchmarks and enables users to define custom benchmarks using a wide variety of parameters, regarding the workload, network settings and blockchain-specific settings. Also, it allows metric collection on the client side to be suitable for an end-to-end evaluation approach.

The paper is further structured as follows. Section 2 gives a brief introduction to the benchmarked blockchain systems. Section 3 gives an explanation of the new benchmarking framework COCONUT, which is used for the benchmarks. Section 4 describes the evaluation environment and the configuration of the benchmarks. The results of the benchmarks conducted, and our insights are discussed in Section 5. Section 6 gives a summary of the lessons learned. Section 7 discusses related work. Section 8 provides a summary and an outlook for further research.

## 2 BLOCKCHAIN SYSTEMS

Blockchain technology is characterized by a data structure composed of blocks, with each block linked to its predecessor. The blocks contain transactions. To ensure tamper protection, the blocks are linked by cryptographic methods, for example with hashes of the predecessor block in the header of each block.

We provide an overview of the examined blockchain systems in Table 2, highlighting key features including their supported consensus algorithms, versions, and transaction structures. Given the range of terms used across technologies, such as "chaincode" or "operation", we will adopt interface execution layer (IEL) as a standardized term for the smart contract constructs.

Blockchain nodes participating in the network must reach a consensus on the order and the validity of the transactions within a block and the blocks themselves. To improve the performance, various consensus algorithms have emerged with some examples shown in Table 2. We can mostly observe byzantine fault tolerant (BFT) and Raft-based implementations used by the selected blockchain systems. One exception is BitShares, which uses the Delegated Proof-of-Stake (DPoS) consensus algorithm. DPoS uses witnesses as authorities who are allowed to produce blocks. Whenever a new block is getting finalized, a new round is started, where a witness can sign and finalize a new block. Corda uses notaries to check for conflicting and already used states. Fabric outsources the consensus component to external entities, called orderers.

**Table 2: Overview of evaluated blockchain systems.**

| System | Consensus Algorithm | Execution Layer | Version | Transaction structure |
|---|---|---|---|---|
| Corda OS & Enterprise | Single notary | Flow | 4.8.6 | Multiple input/output states |
| BitShares | DPoS [28] | Operation | 195881c32f | Multiple operations; added to block |
| Fabric | Raft [46] | Chaincode | 2.2.1 | Single transaction |
| Quorum | Istanbul BFT [44] | Smart Contract | d967b695df | Single transaction |
| Sawtooth | PBFT [20] | Transaction Processor | 1.2.6 | Transactions in atomic batch; added to block |
| Diem (prev. Libra) | DiemBFT [13] | Module/Script | 94a8bca0fa | Single transaction |

Beside the consensus algorithms, some of the blockchain systems also use different techniques to form blocks, which usually consist of transactions. BitShares allows multiple operations that result in single state changes to be put into a single transaction. The atomic transaction will get added to a block. Sawtooth allows to add transactions to an atomic batch. The batch will get added to a block. In this context, atomic means that when a single operation within a transaction or a single transaction within a batch fails, the whole transaction or batch will get rejected and not be added to a block.

Corda holds a unique position among blockchain systems. Corda does not use blocks, but relies on the UTXO model (Unspent Transaction Output) [23, 43]. Using the UTXO model, it is possible to add multiple input and output states to a transaction.

In contrast to the UTXO model, Quorum due to using Ethereum as its basis, uses an account model. Fabric and Sawtooth, for example, allow the implementation of both the account and UTXO model in the interface execution layer.

In summary, benchmarking blockchain systems is more challenging than ever, making a comparison difficult. Given the different architectures, consensus algorithms, transaction structures, network components and different interface execution layer implementation procedures discussed in this section highlight the scope and complexity of performance evaluations.

Besides the blockchain system itself, clients submitting transactions primarily want information about the status of a transaction. This information allows them to initiate further processing or, in the case of an issue, renew the processing of transactions. Our work takes this aspect into account using an end-to-end scenario, where the clients collect transaction finalization notifications, which are used for metric calculation.

## 3 COCONUT BENCHMARKING FRAMEWORK

To take a comprehensive approach, we introduce COCONUT. COCONUT is a benchmarking framework that allows, among other things, the adaptation of consensus algorithms, transaction structures, distribution of network components and block finalization settings of the respective systems for the benchmarks.

We develop COCONUT with the aim to support the following aspects: The extensibility with further systems and interface execution layers, the possibility to adjust a high number of configuration parameters per system, the collection of logs and metrics and the open source availability. The blockchain system represents the entire underlying infrastructure of a blockchain technology under test. The client runs the workloads, collects the data to be evaluated and contains the blockchain access layer (BAL). The blockchain access layer provides the driver implementation that ensures proper communication with the blockchain system and the interface execution layer, for example, smart contracts. The database system provides a persistent storage for the collected evaluation data. COCONUT is designed to be event-based. This means that the clients collect notifications. These notifications include, for example, information about when a new block is added and sent out by the blockchain system. This approach provides an end-to-end scenario in our benchmarks to determine the coherent performance of the blockchain systems analyzed. We define an end-to-end scenario in such a way that a client is involved in the entire process from sending a transaction to receiving the confirmation of a transaction and provides the collected data and metrics for evaluation.

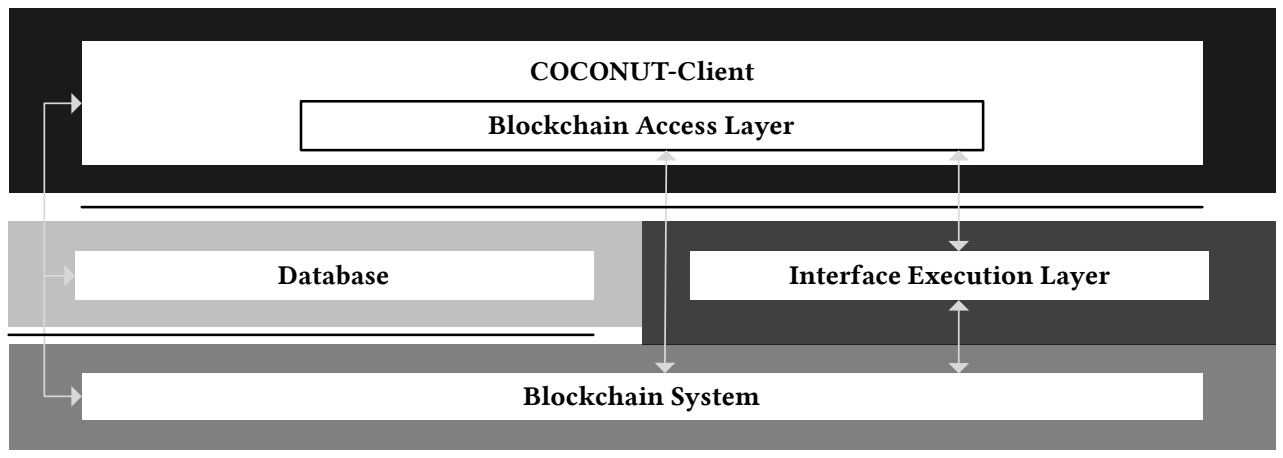

**Figure 1: The base architecture of COCONUT.**

COCONUT was developed with extensibility in mind. Our design proved its flexibility in supporting the seven diverse blockchains we chose.

# 4 EXPERIMENTAL SETTINGS

This section describes all relevant configurations that can be adjusted before the benchmarks are carried out. Our goal is to obtain an assessment of the performance and limitations of the blockchain system by generating a load that aims for a broad comparison that includes an extensive range of settings.

## 4.1 Interface Execution Layers

To run the benchmarks, we use the three different interface execution layer (IEL) implementations listed in Table 3. The first two implementations are inspired by the work of Dinh et al. [25] who characterise and benchmark layers of different systems.

**Table 3: Interface Execution Layers used for the benchmarks.**

| Name | Function(s) | Brief description |
|---|---|---|
| DoNothing | DoNothing | Represents an empty function. |
| KeyValue | Set | Writes a KeyValue pair. |
| | Get | Reads a value based on its key. |
| BankingApp | CreateAccount | Creates checking and saving accounts with defined money. |
| | SendPayment | Sends a payment. |
| | Balance | Checks an account balance. |

The clients send payloads packaged in transactions to the interface execution layer to interact with the blockchain system. This interaction includes, for example, writing and reading data. The interface execution layers consist of functions. The benchmarks that we discuss below are implemented by the interface execution layer. We define a benchmark as the combination of a client sending payloads, which is the workload and the interface execution layer receiving those transactions.

The `DoNothing` IEL is used to evaluate the performance without the execution complexity of the interface execution layer. Therefore, using this benchmark will reveal the performance of the other involved components, for example, the consensus component of the system. We expect the `DoNothing` benchmark to give the best results with respect to throughput compared to the other benchmarks due to its design.

The `KeyValue` IEL allows read and write operations of data, which are essential when dealing with blockchain systems and targets the storage component. The related benchmark is designed in such a way that no duplicates occur during writing.

The `BankingApp` IEL is designed in such a way that side effects occur. This applies in particular to the `SendPayment` function. `SendPayment` sends a payment from $account_n$ to $account_{n+1}$. This should result in overwriting transactions being included in a block, or consuming the same states in the case of Corda. The `SendPayment` function should disclose whether the data in blocks can influence or overwrite each other and therefore impact the performance. In the case of Corda a notary might reject already spent transaction output.

We consider the benchmarks to be sufficient to compare the performance of the chosen systems, as they represent the basic functionality of all the blockchains we have studied. Although the benchmarks are simple, they do give an indication of the raw performance, the performance of read and write operations, and the serialisability issues that can be encountered.

All benchmarks form units. This means that a performed `KeyValue-Set` benchmark is followed by a `KeyValue-Get` benchmark. The same applies to the `BankingApp` benchmark. The sequence is as follows: `BankingApp-CreateAccount`, `BankingApp-SendPayment`, `BankingApp-Balance`.

The execution of the benchmarks is designed so that the system including the interface execution layer is re-provisioned after the benchmark units have been executed and consequently a freshly deployed system is present. This is for example necessary if the system ceased operation during the previously run benchmark unit. The clients are re-provisioned each time a benchmark is run.

## 4.2 Server and network configuration

To ensure a controlled and consistent environment for evaluation, our methodology involved the use of six dedicated servers, all of which are located within our provider's data center in Helsinki, Finland. The decision to confine our evaluation to a single geographical region was deliberate, rooted in our objective to maintain controlled experimental settings, such as the emulated settings. This ensures that result variations arise only from the experimental parameters, free from influences of external geographical disparities.

The servers we employed for our tests possess the following configuration: Processor: AMD Ryzen™7 3700X, Memory: 64 GB DDR4 ECC, Storage: 2 x 1 TB NVMe SoftRAID, Uplink: 1 Gbit/s, Operating System: Ubuntu 18.04. To create a reproducible environment, the servers have sufficient power to ensure proper operation of the blockchain systems. We use Docker to offer an isolated environment for our benchmarking setting. Two of the servers exclusively start instances of the COCONUT client. The remaining four servers provide nodes of the system and the associated components. Table 4 lists the network configuration of the systems. This setting represents a small-scale deployment where only few organizations collaborate using a shared blockchain. However, as our findings will show, many blockchain systems already face performance issues at such a small deployment. Separate scalability experiments for selected systems are performed in Section 5.8.2.

## 4.3 Workload configuration

The COCONUT client starts four concurrent client threads within Docker containers, of which each client thread starts four concurrent workload threads. Each server starts two COCONUT clients. All COCONUT client applications wait for the preparation of each other before executing the workloads in order to create a load distribution that is as uniform as possible. Each of the four COCONUT client applications sends transactions to a different server.

The workload-threads of each COCONUT client application send transactions sequentially, but without waiting for a finalization confirmation, for a period of 300 seconds. We chose this time to allow the benchmarks to run long enough to produce reliable results. The COCONUT client terminates listening on events after 330 seconds. These 30 seconds of additional run-on time, after the sending process, ensure the receipt of delayed finalization confirmations. After 420 seconds, the COCONUT clients finally terminate themselves. We set this time to be long enough to ensure a correct termination of the clients.

**Table 4: Blockchain Systems network configuration.**

| System | # Nodes (n) | Additional components |
|---|---|---|
| Corda OS & Enterprise | 4 | 4 notaries (one per server) |
| BitShares | 4 | 3 witnesses (n - 1) |
| Fabric | 4 | 3 orderers (servers 1-3) |
| Quorum | 4 | None |
| Sawtooth | 4 | None |
| Diem | 4 | None |

These times are guidelines and cannot be completely adhered to in every scenario in the context of each individual benchmark. An example of non-compliance with the specified times are transactions sent that cannot yet be processed on the client or server side. This problem is seen, for example, in systems that cannot process the applied load in a timely manner and temporarily or permanently stop service. To assess this non-trivial problem, we include the duration of the benchmarks in our evaluations. The duration of the benchmark provides information on whether a system can actually process the workload within the 300 seconds in which the clients transmit transactions.

After terminating the COCONUT clients, the collected data is persisted, and the benchmark is considered complete.

## 4.4 System parameter settings

In our work, we focus on the evaluation of two parameters. The first parameter is the maximum number of payloads, wrapped into transactions and batches, to be sent by each COCONUT client per second, which is called the rate limiter (RL) and is equal to the load. The second parameter is the block size or the time after which a block is finalized. We prefer the block size over the time of finalization, provided that the chosen system supports both parameters. Although we prioritize these two parameters, we have evaluated several others. Additional parameters evaluated include the number of batches for Sawtooth, the number of operations for BitShares and the number of accounts needed for Diem. COCONUT offers a vast array of adjustable parameters for further benchmarks.

Fabric and Diem allow for the modification of the block finalization size, as shown in Table 5. BitShares, Quorum and Sawtooth allow for the modification of the block finalization time as shown in Table 6. Some configuration options which could limit or interrupt the execution of our benchmarks are documented in our configuration files.[1] We adjust these configuration options based on our evaluations, but also make sure that they do not affect our benchmarks. An example for such a configuration option is the maximum transaction size in BitShares. A too small setting would prevent large transactions from being processed.

[1]https://www.coconut.sh

All four COCONUT clients use the following rate limiters: 50, 100, 200 and 400 for the systems BitShares, Fabric, Quorum, Sawtooth and Diem. The minimum rate limiter value of 50 per COCONUT client is an empirical value resulting from experiments with the systems Sawtooth and Diem. The maximum rate limiter value of 400 per COCONUT client is based on our experience in dealing with the system Fabric, which in our scenarios, where a single operation/action is used in a transaction, can handle the highest load. For both Corda versions, we apply one tenth of the rate limiters, which corresponds to the values 5, 10, 20 and 40, as both versions cannot handle a higher load.

We use lower values for *max_block_size* than the default value of 3,000 as most of our evaluations resulted in the blocks already not getting fully saturated at our maximum *max_block_size* of 2,000.

We run our BitShares benchmarks with 1, 50 and 100 operations in a transaction. The transaction structures of BitShares and Sawtooth influenced by the used settings are explained in Table 2. Additionally, we run our Sawtooth benchmarks with 1, 50 and 100 transactions within a batch. Corda has no specific parameters in the context of the block size or the time after which a block is finalized in both versions due to the lack of a block structure. For Corda, we use the same configuration for both the open-source version and the enterprise version for reasons of direct comparability.

**Table 5: Block finalization size configuration.**

| System | Parameter | Default | Used |
|---|---|---|---|
| Fabric | MaxMessageCount [32] (MM) | 500 | 100, 500, 1,000, 2,000 |
| Diem | max_block_size [31] (BS) | 3,000 | 100, 500, 1,000, 2,000 |

**Table 6: Block finalization time configuration.**

| System | Parameter | Default | Used |
|---|---|---|---|
| BitShares | block_interval [30] (BI) | 5 s | {1, 2, 5, 10} s |
| Quorum | istanbul.blockperiod [33, 34] (BP) | 1 s | {1, 2, 5, 10} s |
| Sawtooth | sawtooth.consensus.pbft.block_publishing_delay[35] (PD) | 1 s | {1, 2, 5, 10} s |

Some of the systems require a certain time after starting in order to stabilize before they can actually process workloads. We set this time to 180 seconds for BitShares and Quorum, and 60 seconds for Sawtooth, respectively. For the other systems, COCONUT starts with serving the workloads immediately after start up.

## 4.5 Evaluation metrics

In most comprehensive blockchain benchmarking papers ( [45], [25, 38], [51]), the performance evaluation is based on the block generation time in the blockchain. Whether the blockchain application receives confirmation of transaction execution is not further investigated. Whilst data may be persisted in the blockchain, the application sending the data may never be updated about the status of the transaction. After sending a transaction request to the Blockchain, an application usually must wait for a confirmation to complete the processing.

Therefore, we measure the end-to-end transaction time, also known as finalization latency in seconds (FLS), which is important for real-world applications. FLS is calculated from the difference between confirmation time and request time, which are measured exclusively in the COCONUT client simulating a blockchain application executing transactions on a blockchain. As shown in Figure 2, timestamps are obtained just before a transaction request is sent (starttime) and just after the transaction confirmation arrives (endtime). *T1* is the time at which the blockchain receives a request. *T2* is the time at which a transaction is distributed to all nodes and committed, triggering an event notification. These times include e.g. the transaction processing, IEL processing, block formation, validation, consensus and block distribution. These values can be influenced by scaling the network or additional network latency as analyzed in Section 5.8.1. The round trip time (RTT) is the time between the submission of the request and the arrival of the commit. *T0* and *T3* are the important times we use in our end-to-end scenario for metric calculation. It is important to note that in our measurements, a transaction is not considered complete until the transaction has been persisted in all participating blockchain nodes, i.e., a block containing the transaction has been distributed to and persisted in all nodes. After receiving a confirmation message, the COCONUT client, i.e., the blockchain application, can be sure that the transaction has been fully executed and persisted and continue the ongoing processing.

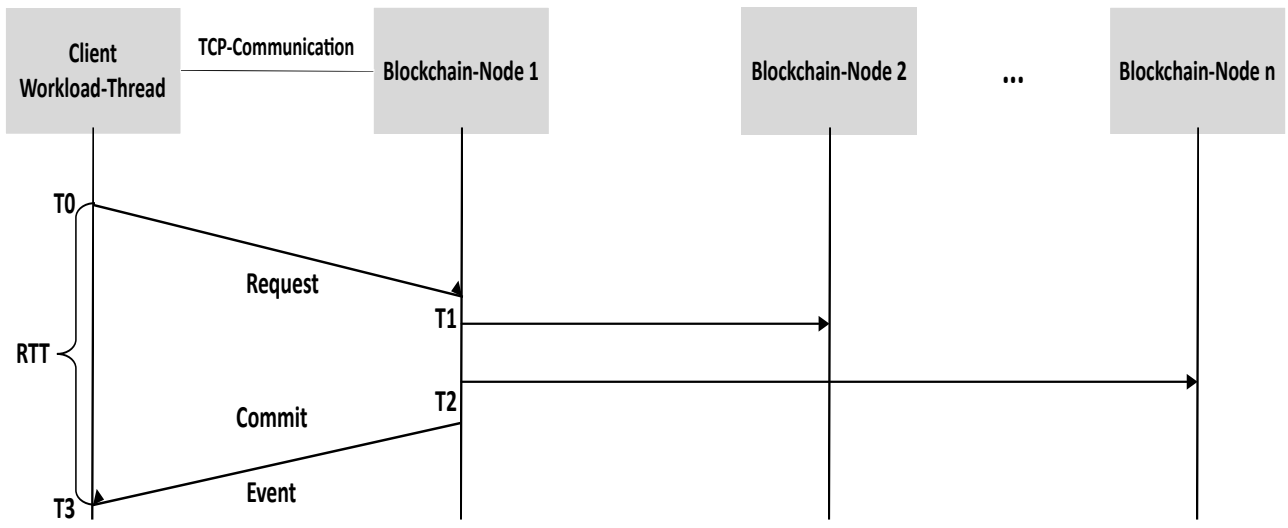

**Figure 2: Visualization of the relevant times for the metric calculation.**

For comparison purposes, we calculate the average finalization latency, which we refer to as MFLS, over all executed transactions using Formula (1). Besides the MFLS, we calculate the throughput measured in transactions per second to assess the performance of the chosen systems. We use Formula (2) to calculate the mean transactions per second (MTPS).

$$MFLS = \frac{1}{r}\sum_{i=1}^{r}\left(\frac{\sum_{j=1}^{t}\left(\text{endtime}_j - \text{starttime}_j\right)}{t}\right)_i \tag{1}$$

Here, $r$ corresponds to the number of repetitions of the benchmarks, $t$ reflects the number of transactions, $endtime_j$ is the time at which a client receives confirmation of a finalized transaction and $starttime_j$ is the time shortly before the transaction is sent.

$$MTPS = \frac{1}{r}\sum_{i=1}^{r}\left(\frac{t}{t_{\text{lrtx}} - t_{\text{fstx}}}\right)_i \tag{2}$$

Here, $t_{lrtx}$ is the reception time of the last transaction received across all COCONUT clients, $t_{fstx}$ corresponds to the time just before the first transaction is sent, across all COCONUT clients.

$t_{fstx}$ is measured from the client that sends the first and $t_{lrtx}$ from the client that receives the last transaction across all clients. Additionally, clients only receive the confirmation when the transaction is finalized on all blockchain nodes.

Also, we include the duration of the benchmark in our evaluation. This value is important to be able to evaluate if the system stops processing transactions at an earlier time than the originally expected end of the benchmark or violates the liveness criteria by ceasing operation.

$$Duration\ (D) = t_{\text{lrtx}} - t_{\text{fstx}} \tag{3}$$

In addition to the duration of the benchmark, we also analyse the number of expected, received and not received transactions (NoT). All values are calculated as an average across all clients.

BitShares is an exception in the context of the MTPS calculation due to the scenarios with multiple operations within one transaction. We treat each operation the clients receive as a single transaction.

# 5 RESULTS

This section discusses our key findings. Our results are comprehensive, spanning seven different systems. The modification and evaluation of various parameter settings, such as the number of transactions per block, further validate the depth of our analysis. Figure 3 underpins the best throughput and corresponding finalization latency values per benchmark and system in the form of a heat map, which allows a direct performance comparison across the seven analyzed systems. When discussing specific anomalies of the evaluated systems, we additionally present tables that include the standard deviation (SD), the standard error of the mean (SEM), and the 95% confidence interval (95% CI). Finally, we discuss the performance of all benchmarks in the context of the different systems. This facilitates a performance comparison among the systems. We make all necessary configurations to reproduce the performed benchmarks as well as all results themselves available online.[2]

## 5.1 Corda OS

Corda OS proves to be the weakest system in terms of performance in our benchmarks. We list the results of the `KeyValue-Set` benchmark in Table 7 and Table 8. The `KeyValue-Set` benchmark is to be compared with the `KeyValue-Get` benchmark, as both iteratively check whether a KeyValue pair exists. For the `KeyValue-Get` benchmark, all transactions fail and are not received by the clients for all loads (RL = 20 to RL = 160) without exception for the reasons listed below. The reasons for the low performance include the following:

(1) We only use the functions offered by Corda for the implementation of the interface execution layer. These functions require, for example in the case of a read operation, iterating over each KeyValue pair to find a specific one. This greatly slows down the processing of transactions. An alternative to using these functions would be to use native queries targeting the underlying database system of Corda OS, the H2 in-memory database [2]. The implementation of native queries is a potential workaround that is not part of Corda itself. In this respect, we are not pursuing this approach any further.
(2) In our scenarios, each of the four nodes must sign the submitted transaction. The node that receives the transaction first forwards the transaction to the other three nodes in the network. Corda OS does this serially.

[2] https://www.coconut.sh

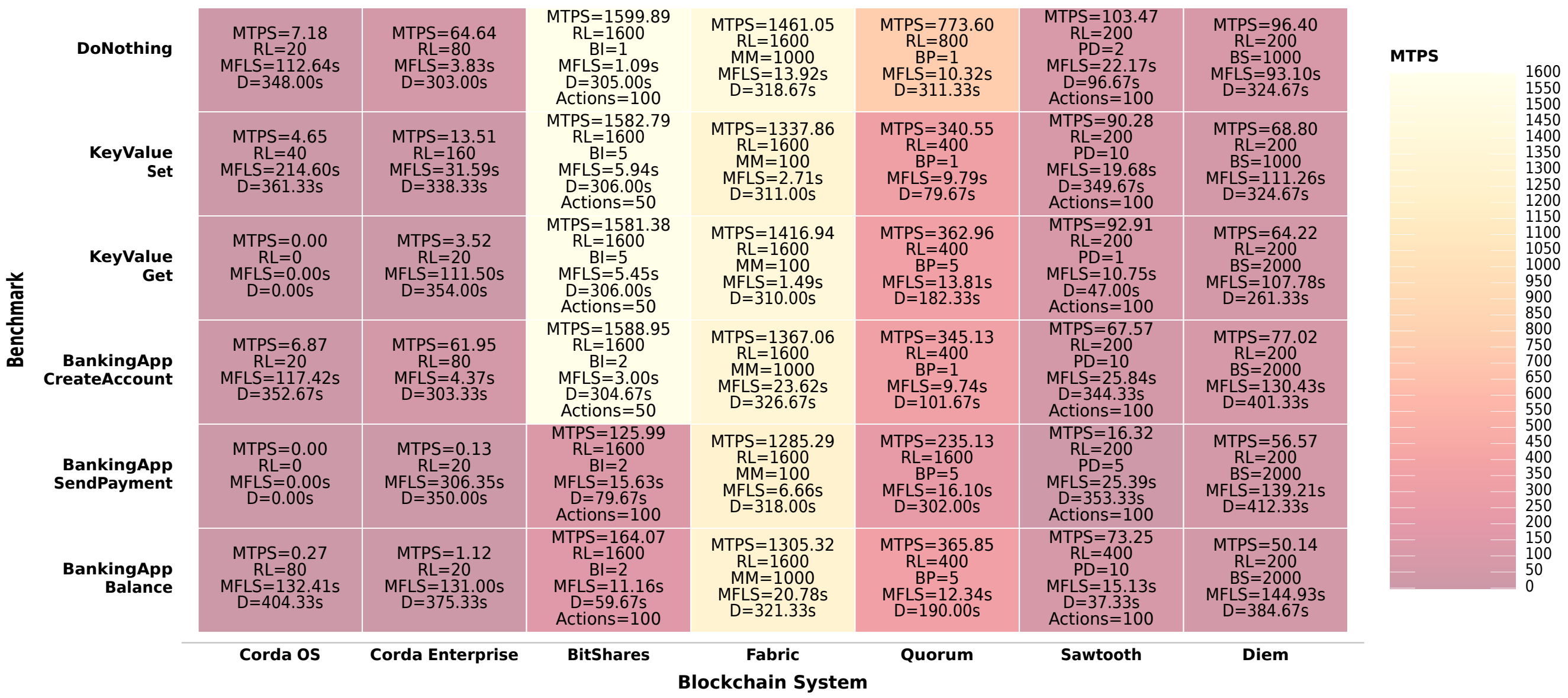

**Figure 3: Best MTPS values with corresponding MFLS and Duration for the shown configurations.**

| RL | MTPS | SD | SEM | 95% CI | MFLS | SD | SEM | 95% CI |
|---|---|---|---|---|---|---|---|---|
| 20 | 4.08 | 0.01 | 0.01 | ±0.03 | 151.93 | 0.70 | 0.40 | ±1.75 |
| 160 | 1.04 | 0.16 | 0.09 | ±0.41 | 227.39 | 7.29 | 4.21 | ±18.1 |

**Table 7: MTPS and MFLS for Corda OS - KeyValue-Set**

| RL | Received NoT | Expected NoT | SD | SEM | 95% CI |
|---|---|---|---|---|---|
| 20 | 1439.00 | 6000.00 | 7.94 | 4.58 | ±19.72 |
| 160 | 374.33 | 48000.00 | 62.94 | 36.34 | ±156.35 |

**Table 8: # of Transaction values for Corda OS - KeyValue-Set**

Corda OS has MTPS values ranging from 0.27 to 7.18. The latency of finalization increases with increasing load. In addition, the number of failed transactions increases.

## 5.2 Corda Enterprise

In contrast to Corda OS, Corda Enterprise achieves better results in all scenarios. For the comparison between Corda Enterprise and Corda OS, we choose to depict the results of the `KeyValue-Set` benchmark in Table 9 and Table 10. The displayed values in the range of RL = 20 to RL = 160 show hardly any differences. It can also be seen that the best results are achieved with the `BankingApp-CreateAccount` and `DoNothing` benchmarks which do not read any data. The fact that benchmarks that do not read data perform better than those that do read data is already explained by the enumeration for Corda above. Benchmarks that only write data show higher throughput, a lower finalization latency and a lower number of lost transactions. In the scenarios involving read operations, a decrease in performance can be seen with an associated reduction in throughput, a drop in received transactions and an increase in finalization latency. The use of multithreading and parallel signing of transactions by the nodes in the network improve performance as compared to Corda OS [48]. No benchmark fails completely, but the range from 0.13 to 64.64 MTPS is hardly comparable to the results achieved by the other systems.

| RL | MTPS | SD | SEM | 95% CI | MFLS | SD | SEM | 95% CI |
|---|---|---|---|---|---|---|---|---|
| 20 | 12.84 | 0.15 | 0.09 | ±0.38 | 22.81 | 1.89 | 1.09 | ±4.7 |
| 160 | 13.51 | 0.16 | 0.09 | ±0.4 | 31.59 | 2.83 | 1.64 | ±7.05 |

**Table 9: MTPS and MFLS for Corda Enterprise - KeyValue-Set**

| RL | Received NoT | Expected NoT | SD | SEM | 95% CI |
|---|---|---|---|---|---|
| 20 | 4249.67 | 6000.00 | 34.49 | 19.91 | ±85.67 |
| 160 | 4571.00 | 48000.00 | 68.61 | 39.61 | ±170.43 |

**Table 10: # of Transaction values for Corda Enterprise - KeyValue-Set**

## 5.3 BitShares

BitShares delivers different results depending on the benchmark and parameter settings. The results of the `DoNothing` benchmark displayed in Table 11 and Table 12 show that at a load of RL = 1,600, the throughput is higher than with `Fabric`, and the COCONUT clients receive all transactions correctly. The finalization latency is close to the specified *block_interval*, indicating that the processing of transactions is timed to match the *block_interval.* In the `BankingApp-SendPayment` benchmark, which is intended to show whether there are problems with the serial processing of transactions, and the consecutive `BankingApp-Balance` benchmark, we record almost exclusively lost transactions. Due to the atomicity of transactions, if an operation fails, the whole transaction is discarded. This suggests that BitShares does not include interacting operations or transactions in a block, resulting in reduced throughput and increased finalization latency, which increases as the load increases. In addition, when operations interfere with each other during execution, the experiment duration is shorter than the 300 seconds of the transaction sending process. This indicates that either the system is no longer sending out finalized transactions, which consequently violates the liveness criterion, or the processing takes longer than the execution of our benchmark. The concept of an operation is defined in Table 2 and is comparable with the concept of batches with Sawtooth. The throughput varies between 125.99 and 1,599.89 MTPS according to our measurements. The block statistics show in scenarios where the finalization latency does not match the *block_interval*, whether the witnesses still generate the blocks correctly. With only one operation per transaction, the minimum MTPS value of 59.01 is obtained for the `BankingApp-SendPayment` benchmark. For the `DoNothing` benchmark, the maximum MTPS values of up to 589.58 can be observed when using one operation per transaction.

| RL | BI | MTPS | SD | SEM | 95% CI | MFLS | SD | SEM | 95% CI |
|---|---|---|---|---|---|---|---|---|---|
| 1600 | 1s | 1599.89 | 0.19 | 0.11 | ±0.47 | 1.09 | 0.09 | 0.05 | ±0.22 |

**Table 11: MTPS and MFLS for BitShares - DoNothing. (BI = block_interval)**

| RL | BI | Received NoT | Expected NoT | SD | SEM | 95% CI |
|---|---|---|---|---|---|---|
| 1600 | 1s | 487966.67 | 480000.00 | 57.74 | 33.33 | ±143.42 |

**Table 12: # of Transaction values for BitShares - DoNothing. (BI = block_interval)**

## 5.4 Fabric

Fabric shows consistent performance across all benchmarks. We discuss the behaviour of the `BankingApp-SendPayment` benchmark shown in Table 13 and Table 14 , as it differs from the other systems. Fabric appends every processed transaction to the blockchain, even those transactions not carried over to the *world state*, for example, due to serializability issues. The *world state* provides a consistent view of the data and represents a subset of the blockchain data. We include every transaction appended to the blockchain in our results. The blocks can be saturated up to the maximum setting *MaxMessageCount* (MM) = 2,000. The throughput reaches a maximum of 1,285.29 to 1,461.05 MTPS, whereby the modification of the *MaxMessageCount* value does not reveal a high impact here. Furthermore, at the maximum load (RL = 1,600), not all expected transactions can be received, so that some are lost. This behaviour can be observed across all scenarios. We only observe this behaviour with the choice of the consensus algorithm Raft, but not when using Apache Kafka [17]. Apache Kafka produces overhead due to its architecture, which leads to slower processing of the transactions, but is much more mature than Raft in its development. We attribute the failing transactions and malfunctioning orderers to this circumstance. Overall, Fabric has no other special features and shows superior MTPS-related results compared to the other systems in most scenarios. Clients constantly receive a block-related event every second, which corresponds to the expected runtime. This implies that `Fabric` can handle the applied load without negative influences, like severe processing delays or lost transactions from the client perspective.

| RL | MM | MTPS | SD | SEM | 95% CI | MFLS | SD | SEM | 95% CI |
|---|---|---|---|---|---|---|---|---|---|
| 800 | 100 | 801.36 | 1.53 | 0.88 | ±3.8 | 0.22 | 0.01 | 0.01 | ±0.03 |
| 1600 | 100 | 1285.29 | 64.39 | 37.18 | ±159.96 | 6.66 | 4.74 | 2.74 | ±11.78 |

**Table 13: MTPS and MFLS for Fabric - BankingApp-SendPayment. (MM = MaxMessageCount)**

| RL | MM | Received NoT | Expected NoT | SD | SEM | 95% CI |
|---|---|---|---|---|---|---|
| 800 | 100 | 240140.67 | 240000.00 | 9.81 | 5.67 | ±24.38 |
| 1600 | 100 | 408749.00 | 480000.00 | 23837.96 | 13762.85 | ±59216.78 |

**Table 14: # of Transaction values for Fabric - BankingApp-SendPayment. (MM = MaxMessageCount)**

## 5.5 Quorum

For Quorum, we highlight the results of the `BankingApp-Balance` benchmark in Table 15 and Table 16. We select this benchmark as

| RL | BP | MTPS | SD | SEM | 95% CI | MFLS | SD | SEM | 95% CI |
|---|---|---|---|---|---|---|---|---|---|
| 400 | 2s | 0.00 | 0.00 | 0.00 | ±0 | 0.00 | 0.00 | 0.00 | ±0 |
| 400 | 5s | 365.85 | 10.63 | 6.14 | ±26.42 | 12.34 | 0.67 | 0.38 | ±1.66 |

**Table 15: MTPS and MFLS for Quorum - BankingApp-Balance. (BP = istanbul.blockperiod)**

| RL | BP | Received NoT | Expected NoT | SD | SEM | 95% CI |
|---|---|---|---|---|---|---|
| 400 | 2s | 0.00 | 120000.00 | 0.00 | 0.00 | ±0 |
| 400 | 5s | 69476.33 | 120000.00 | 456.37 | 263.49 | ±1133.69 |

**Table 16: # of Transaction values for Quorum - BankingApp-Balance. (BP = istanbul.blockperiod)**

Quorum, unlike BitShares, does not show significant performance degradation on the BankingAppSendPayment benchmark and the consecutive BankingApp-Balance benchmark. Quorum shows a comparable performance with BitShares using the BankingApp-Balance benchmark, when using only single operations per transaction. We attribute this to the better prevention of forks with the used consensus algorithm and Ethereum's order-execute paradigm [26]; since Quorum is an extension of Ethereum, it implements the same paradigm. For an *istanbul.blockperiod* ≤ 2, Quorum shows a significant problem. We can see that all transactions fail, and the Quorum nodes generate empty blocks. We notice that when *istanbul.blockperiod* is low, combined with a high rate limiter value, Quorum adds transactions to a queue, but the queue is no longer processed. We also observe this behaviour when using the Raft consensus algorithm. This suggests a strong violation of the liveness criterion. To avoid the problem, we recommend either an increase of the *istanbul.blockperiod* value or a decrease of transaction rate. With *istanbul.blockperiod* (BP) = 5 and RL = 400, 365.85 MTPS are reached with a MFLS of 12.34 s. Besides the DoNothing benchmark, which shows MTPS of 773.60 and an associated MFLS of 10.32 s, the other benchmarks show MTPS values from 235.13 to 365.85 with associated MFLS values from 9.74 s to 16.10 s.

## 5.6 Sawtooth

The results of the BankingApp-CreateAccount benchmark shown in Table 17 and Table 18 are particularly striking in Sawtooth. The results show a pattern that we also observe with the DoNothing benchmark and the KeyValue-Set benchmark. As the load increases, there is a decrease in throughput with an increase in finalization latency. This is accompanied by an increase in the number of lost transactions. With Sawtooth, one factor is especially decisive for the non-reception of expected transactions. This factor is the management of a queue that rejects new incoming transactions if the occupancy of the queue is too high. In this case, it is required to re-send the rejected transaction or the atomic batch. The concept of a batch is comparable with the concept of operations with Bitshares as multiple transactions can be added to a single atomic batch. The high number of lost transactions is evident across all benchmarks performed and is mainly due to the queue being too busy. Furthermore, we can observe that adjusting the *sawtooth.consensus.pbft.block_publishing_delay* value does not reveal any significant difference. The blocks cannot be saturated in any scenario. In each scenario, we achieve optimal values in the range of 16.32 MTPS for the BankingApp-SendPayment benchmark and up to 103.47 MTPS for the DoNothing benchmark. The corresponding MFLS values for all benchmarks range from 10.75 s to 25.84 s. Moreover, the high number of failed transactions in the case of the BankingApp-SendPayment benchmark is due to the atomic treatment of batches. All MTPS-related optimal values are achieved with 100 transactions per batch, which explains the lower number of received transactions and consequently the lower throughput compared to the other benchmarks shown in Figure 3 for Sawtooth. This can be explained by the rejection of batches when the queue is full and the atomic structure of batches. If a transaction fails within a batch, the entire batch of 100 transactions is completely discarded. Using one transaction per batch, the best MTPS values range from 26.14 to 35.13 with an associated MFLS in the range of 9.37 s to 13.72 s. The KeyValue-Get, BankingApp-SendPayment and BankingApp-Balance benchmarks fail in every case when using a single transaction per batch.

| RL | PD | MTPS | SD | SEM | 95% CI | MFLS | SD | SEM | 95% CI |
|---|---|---|---|---|---|---|---|---|---|
| 200 | 1s | 66.70 | 0.34 | 0.20 | ±0.84 | 26.40 | 0.02 | 0.01 | ±0.05 |
| 1600 | 1s | 14.27 | 3.23 | 1.86 | ±8.02 | 238.45 | 21.12 | 12.19 | ±52.45 |
| 200 | 10s | 67.57 | 1.32 | 0.76 | ±3.28 | 25.84 | 0.72 | 0.41 | ±1.78 |
| 1600 | 10s | 15.65 | 0.52 | 0.30 | ±1.31 | 225.73 | 3.45 | 1.99 | ±8.57 |

**Table 17: MTPS and MFLS for Sawtooth - BankingApp-CreateAccount. (PD = block_publishing_delay)**

| RL | PD | Received NoT | Expected NoT | SD | SEM | 95% CI |
|---|---|---|---|---|---|---|
| 200 | 1s | 23033.33 | 60000.00 | 57.74 | 33.33 | ±143.43 |
| 1600 | 1s | 4666.67 | 480000.00 | 1101.51 | 635.96 | ±2736.31 |
| 200 | 10s | 23266.67 | 60000.00 | 378.59 | 218.58 | ±940.48 |
| 1600 | 10s | 5133.33 | 480000.00 | 115.47 | 66.67 | ±286.85 |

**Table 18: # of Transaction values for Sawtooth - BankingApp-CreateAccount. (PD = block_publishing_delay)**

## 5.7 Diem

To characterize the performance of Diem, we select the KeyValue-Get benchmark displayed in Table 19 and Table 20. We set the lower limit of *max_block_size* to 100 and the upper limit to 2,000. The tables show that there are differences between the minimum and maximum limits, but these have only a minor impact on the overall performance. Much like Sawtooth, a significant number of transactions fail, adversely affecting the overall performance. As rate limiter values increase, throughput decreases, whereas finalization latency increases. The MTPS values range from 50.14 to 96.40 with the corresponding MFLS values ranging from 93.10 s to 144.93 s. The best MTPS-related values are obtained with RL = 200 and *max_block_size* (BS) ≥ 1,000. Although the specified *max_block_size* can be approximately saturated in most scenarios, we can see that not all transactions are processed. Balster [40] describes various performance problems, including the "spiking" behaviour of Diem, in which validators temporarily stop validating further transactions. We can observe similar behaviour during the execution of the Diem benchmarks which leads to the blocks not getting fully saturated.

| RL | BS | MTPS | SD | SEM | 95% CI | MFLS | SD | SEM | 95% CI |
|---|---|---|---|---|---|---|---|---|---|
| 200 | 100 | 38.32 | 10.55 | 6.09 | ±26.22 | 67.97 | 4.65 | 2.69 | ±11.56 |
| 1600 | 100 | 11.83 | 1.45 | 0.84 | ±3.59 | 81.30 | 1.67 | 0.97 | ±4.15 |
| 200 | 2000 | 64.22 | 2.57 | 1.48 | ±6.39 | 107.78 | 2.75 | 1.59 | ±6.83 |
| 1600 | 2000 | 36.65 | 5.26 | 3.04 | ±13.07 | 150.35 | 13.67 | 7.89 | ±33.97 |

**Table 19: MTPS and MFLS with statistics for Diem - KeyValue-Get. (BS = max_block_size)**

| RL | BS | Received NoT | Expected NoT | SD | SEM | 95% CI |
|---|---|---|---|---|---|---|
| 200 | 100 | 7365.33 | 60000.00 | 603.00 | 348.14 | ±1497.93 |
| 1600 | 100 | 3887.67 | 480000.00 | 470.47 | 271.62 | ±1168.7 |
| 200 | 2000 | 16752.67 | 60000.00 | 699.99 | 404.14 | ±1738.88 |
| 1600 | 2000 | 11172.67 | 480000.00 | 1238.53 | 715.06 | ±3076.67 |

**Table 20: # of Transaction values with statistics for Diem - KeyValue-Get. (BS = max_block_size)**

## 5.8 Impact of network latency and scalability analysis

In this section, we analyse the impact of network latency and scalability scenarios, with 8, 16 and 32 nodes.

*5.8.1 Impact of network latency.* We emulate the network latency in a scenario where the servers are equidistant to analyse their impact on the performance of each system. To emulate the latency, we use netem [9] with normal distribution parameters $\mu$ = 12 ms, $\sigma^2$ = 2 ms. The $\mu$-value is derived based on European servers provided by WonderNetwork [54]. As highlighted in Section 4.2, one of our primary objectives was to achieve consistent and repeatable results. In the context of network latency, this consistency was ensured using netem, which provides a robust mechanism to introduce and control latency systematically. By doing so, we aimed to negate the effects of external fluctuations and influences, which are often associated with unpredictable network conditions. Such uncontrollable fluctuations could potentially compromise the repeatability and consistency of our experiments.

The heat map in Figure 4 shows the effect of the latency and allows a direct comparison of the achieved performance. These values are based on the parameters used to obtain the best MTPS values shown in Figure 3. They include throughput, finalization latency and the duration of the benchmark.

Corda OS hardly reacts to the network latency. Corda Enterprise also shows little response except for two benchmarks. The BankingApp-SendPayment benchmark is the only benchmark that requires communication with the notary to check whether states have already been consumed. As a result, the BankingApp-SendPayment benchmark is more affected by the emulated latency due to the additional communication with the notary. The BankingApp-SendPayment and the subsequent BankingApp-Balance benchmarks fail completely with emulated latency due to this reason in our scenario.

BitShares shows a diffuse picture with hardly changed results in the DoNothing benchmark and strong performance drops in the KeyValue-Set, KeyValue-Get and the BankingApp-CreateAccount benchmarks. Here, MTPS values in the range of 579.45 to 1,046.87 are achieved compared to MTPS values in the range of 1,581.38 to 1,588.95 without emulated latency.

The BankingApp-SendPayment and BankingApp-Balance benchmarks show MTPS values in the range of 6.62 to 9.96, which corresponds to a fraction of the results of 125.99 to 164.07 achieved without emulated latency. We can explain this behaviour using the durations of 356.00 s and 369.33 s of the experiments. We use these durations to calculate the MTPS values. Having long runtime durations with low throughput results in lower MTPS values. While the DoNothing benchmark achieves to received the number of expected transaction confirmations, this is not the case for the other benchmarks explaining the values shown.

Fabric consistently shows results in the range of 866.30 to 898.78 MTPS, which is a loss of 33% to 40% compared to the results without emulated latency. Fabric is thus sensitive to network latency. The main reason for this loss is the overhead that arises from the additional communication with the orderers who are responsible for reaching consensus.

Quorum achieves similar results in terms of throughput and finalization latency, both with and without emulated latency. Although the DoNothing benchmark achieves lower MTPS-related values with emulated latency of 605.04 MTPS, the values of the other benchmarks in the range of 243.13 to 362.50 MTPS are like the values without emulated latency, which are in the range of 235.13 to 365.85 MTPS.

Sawtooth hardly reacts to the emulated latency. Only the BankingApp-Balance benchmark shows a weaker result. The value without emulated latency is 73.25 MTPS compared to 30.24 MTPS with emulated latency. As with BitShares, this is due to the longer execution time of the benchmark.

Similar to Sawtooth, Diem reacts very indifferently to the emulated latency in our scenarios comparable to the MTPS- and MFLS values without emulated latency.

**Summary:** Corda OS, Quorum, Sawtooth and Diem hardly react to the emulated latency in our experiments. Corda Enterprise, BitShares and Fabric sometimes show performance drops. Using emulated latency, BitShares shows different performance patterns across all benchmarks. While Corda Enterprise has several optimisations compared to Corda OS, the performance degradation of Corda Enterprise and Fabric is due to the increased communication overhead with the additional network components.

*5.8.2 Scalability.* We scale the nodes in the network to analyse their impact on the performance of the selected systems. Figure 5 shows the results of the MTPS values obtained. The peer numbers, grouped by the individual blockchain systems are mapped on the x-axis. The y-axis represents the MTPS values according to a logarithmic scale. The figure shows the results for 8, 16 and 32 nodes. As a basis, we use the same settings as in Section 5.8.1. Instead of six servers, we use ten servers whose configuration corresponds to the description from Section 4.2. We distribute 8, 16 and 32 nodes of the system successively to eight instead of four servers according to a round-robin procedure. Each server starts a maximum of four nodes. We discuss the DoNothing benchmark, which provides the best MTPS in all scenarios.

With an increase in the size of the network, Corda OS shows the MTPS values to fall. With 32 nodes, all DoNothing benchmarks fail. The main reason for this is the serial processing of transactions beside the additional communication with the other nodes.

Corda Enterprise can process transactions in all scaling scenarios. However, a similar trend to Corda OS with decreasing MTPS is emerging here when increasing the size of the network. The main reason here is the additional communication with the other nodes.

BitShares shows only marginal fluctuations in all scenarios and achieves similar MTPS values as with four nodes.

While the results differ only slightly for Fabric with four and eight nodes, it is noticeable that all benchmarks fail in the scenarios with 16 and 32 nodes. After analysing the two failing scenarios, we see that the nodes and the orderers successfully process and finalise the transactions, but the clients do not receive any confirmation. Using an end-to-end approach like measuring the metric data on the blockchain nodes would circumvent this problem, but would also undermine the role of the clients as the ultimate recipient. This clearly shows the difference to other measurements obtained from papers comparing multiple blockchain systems [25, 39, 45].

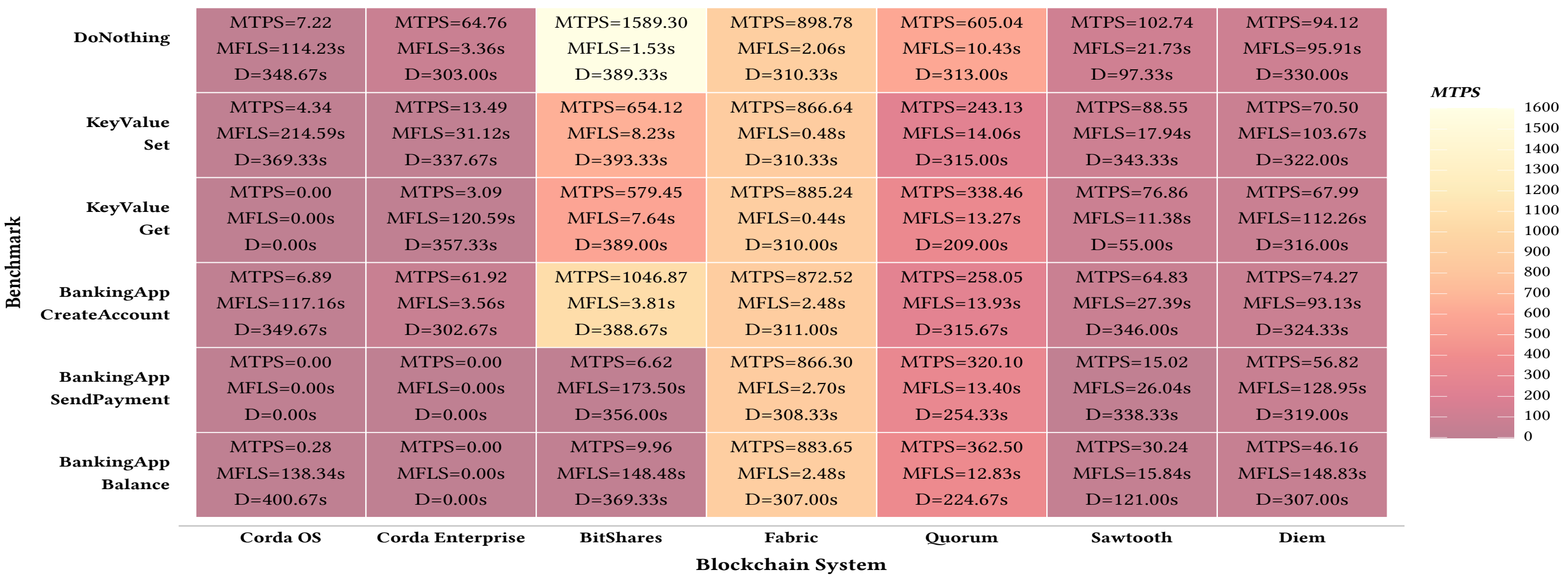

**Figure 4: The best MTPS values displayed are achieved with the configuration values displayed in Figure 3 and the applied emulated latency. Beside the best MTPS values, the corresponding MFLS and Duration are shown.**

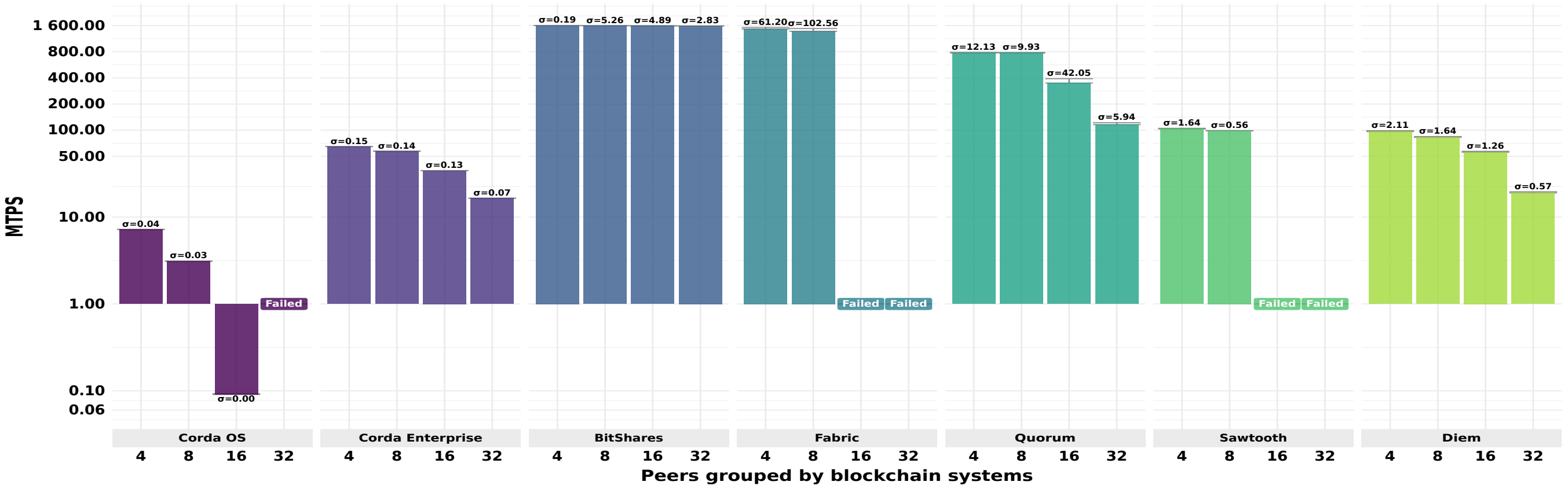

**Figure 5: The influence of network scalability with 8, 16 and 32 nodes on MTPS values based on the DoNothing benchmark.**

Like Corda Enterprise, Quorum shows a downward trend in MTPS values from eight nodes onwards due to the communication overhead of the nodes and the consensus algorithm.

Sawtooth behaves similarly with four nodes and eight nodes. In the scenarios with 16 and 32 nodes, the `DoNothing` benchmarks fail, as with Fabric. In contrast to Fabric, we locate the problem on the side of the nodes since all transactions remain in the pending state without being finalized.

Diem shows a downward trend in MTPS values from 8 nodes onwards. We observe similar behaviour with Corda Enterprise and Quorum. The problem is the communication overhead of the nodes and the consensus algorithm.

**Summary:** Beside BitShares, all other systems show a downward trend in MTPS when we add nodes to the network. This behaviour can also be seen with the vast majority of the other benchmarks. For Diem, Sawtooth and Quorum, the BFT implementation is a reason for the performance drop. With Corda, we let all nodes sign a transaction, which explains the performance drop. For BitShares, the DPoS implementation with shifting witnesses finalizing blocks is a reason for the constant performance.

## 6 LESSONS LEARNED

Here, we summarise the lessons learned based on the observed results.

**Parameter impact:** While the adaptation of the parameters we examined only plays a rather minor role in the systems Fabric, Sawtooth and Diem, BitShares and especially Quorum show advantages of adapting block finalization parameters. However, the respective results are only meaningful in their entirety together which includes the parameters and the interface execution layer used. Our work combines analyzing various blockchain systems, the respective parameters and interface execution layers.

**User Considerations:** For users, one of the crucial takeaways from our study is understanding the transaction, latency, and error rates of each system. Utilizing an end-to-end scenario in our analysis offers a comprehensive view that directly benefits users. Moreover, our analysis indicates that the user perspective reveals different performance results as opposed to measuring directly on the nodes. This knowledge is paramount when selecting a technology, as it can directly impact the efficiency and reliability of the desired application or platform.

**Network latency:** Whilst we did not observe a major impact of network latency with some systems, some showed weaknesses when it comes to communication with other network components.

**Scalability:** We observed that many systems show performance degradation with increasing number of nodes, which is mostly due to their consensus mechanisms. As with the impact of network latency, further research on scalable consensus for blockchains is necessary. In a network that consists of many peers, where only a small subset of nodes need to sign a transaction at a time, Corda could achieve higher performance than Fabric.

**Summary:** In summary, each of the systems studied has its own weaknesses when it comes to factors such as liveness, latency impact, scalability, performance, implementation features, and documentation. Some weaknesses, like liveness, scalability, and performance, can be traced back to specific system components, such as the consensus mechanism used. Nevertheless, certain concepts, like bundling multiple state changes into a single transaction or the externalization of the consensus component, emerge as promising strategies for future improvements.

## 7 RELATED WORK

Our work is not the only research on performance analysis of blockchain systems. Dinh et al. [25] present BlockBench, one of the first tools for performance evaluation of blockchain systems. Due to the differences between the systems they used: Ethereum, Parity and Fabric v0.6.0-preview, our results are not very comparable. For example, Fabric v0.6.0-preview implements the order-execute paradigm [26], while Fabric 2.2.1 follows the execute-order-validate paradigm [16]. In contrast to Dinh et al., we evaluate seven systems with multiple parameter settings, but we adapt some of their ideas to implement custom interface execution layers.

Nasrulin et al. [45] present Gromit, a performance evaluation framework which they use to analyse seven different systems. Like us, they analyse Fabric version 1.4.9, Diem and BitShares, among others. While the obtained evaluation results of BitShares and Fabric are comparable to our results, our evaluations of Diem differ strongly. Nasrulin et al. achieve a throughput in the range of 1,000 TPS, whereas we achieve a maximum of just below 100 TPS in our end-to-end scenario.

Gramoli et al. [39] introduce Diablo, a performance evaluation suite they use to benchmark seven different systems. Gramoli et al. take the approach of implementing and evaluating five workloads based on real-world scenarios. This workloads include stock trading as well as requests to websites at peak times. The conclusion of the work is that no blockchain system can process the load created. While Diablo is comprehensive, no end-to-end scenarios are considered, and no parameter settings investigated; all systems are run with default settings. Also, the metrics are extracted from the logs of the respective systems. This way of recording metrics means that transactions could be completed on the blockchain but not captured by the recipients expecting a confirmation, the clients. Confirmations on the client side can be lost, for example, in case of synchronisation problems of the nodes.

The following two papers go deeper into BitShares-like and Fabric-based systems, while our study is broader. Xu et al. [19] analyse the performance of EOS [27] based on a variety of parameters. The parameters analysed include the transaction size, the number of transactions a client sends, an underlying round-trip time latency, packet loss in the network, the number of clients and the number of block producers. EOS is a system whose database structure is based on BitShares. Here, a maximum result of 250 TPS is achieved in an artificial evaluation environment. In an environment with a round-trip time latency of 50 ms, the number of TPS drops to 50.

Chacko et al. [21] use and analyse various Fabric-based derivatives and Fabric v1.4 to explore why transactions fail and how transaction failures can be mitigated. It becomes clear that block size is a key determinant of transaction failures.

Balster [40] uses a simulation to analyse Diem's core statement, which implies that 1,000 payment transactions can be processed per second on 100 validator nodes. The outcome of Balster's study shows that the results of the simulation at the time of writing do not come close to the value listed above, published by Diem.

Our research uniquely positions itself within the broader landscape of related works. While we introduce a custom framework and evaluate multiple blockchain systems as seen in references like [25, 39, 45], our study stands out due to its concentrated focus on end-to-end scenarios. Though we recognize and appreciate the contributions of prior studies, our analysis suggests that most have not centrally emphasized this aspect. This distinctive approach not only enriches the existing body of research but also sets our work apart. Given our specialized emphasis and methodology, our findings and conclusions might diverge significantly, rendering direct comparisons with other papers somewhat challenging. In light of this, we believe it's crucial to underscore this differentiation within the context of existing literature.

## 8 CONCLUSIONS

The results of our experiments provide an overview of the performance and associated limitations of seven blockchain systems. The analysis of different blockchains in connection with specific interface execution layers, combined with the focus on comparable parameter settings, especially related to block finalisation, is a key feature of our work. Contrary to our initial expectation, parameter adjustments did not significantly impact our measurement results — a key finding of our study. Our end-to-end approach to metric measurement distinguishes our work from other comparable studies.

We will use the results of this work as a guide in our further research to mitigate the weak points and the resulting performance decay.